\documentclass{aa}  
\usepackage{graphicx}
\usepackage{txfonts}
\DeclareMathAlphabet{\mathcal}{OMS}{cmsy}{m}{n} 
\usepackage{keyval}
\usepackage{hyperref}

\usepackage{xcolor}
\begin{document} 

   \title{Dynamical analysis of the 5:1 mean-motion resonance of the HD\,202206 system}

   \author{Yingyi Cao
          \inst{1}
          \and
          Man Hoi Lee\inst{1,2,3}
          \and                                                 
          Sabine Reffert\inst{4}
          \and
          Trifon Trifonov
          \inst{4,5}
          \and
          Stefan Stefanov \inst{5,6}
          }

   \institute{Department of Earth and Planetary Sciences, The
University of Hong Kong, Pokfulam Road, Hong Kong \\
            \email{astrocyy@connect.hku.hk}
         \and
             Department of Physics, The University of Hong Kong, Pokfulam Road, Hong Kong            
        \and
            Hong Kong Institute for Astronomy and Astrophysics, Hong Kong
        \and
              Landessternwarte, Zentrum für Astronomie der Universität Heidelberg, Königstuhl 12, 69117 Heidelberg, Germany
        \and
              Department of Astronomy, Faculty of Physics, Sofia University “St Kliment Ohridski”, 5 James Bourchier Blvd, 1164 Sofia, Bulgaria
        \and
            Institute of Astronomy and National Astronomical Observatory, Bulgarian Academy of Sciences, Tsarigradsko Shose 72, BG-1784 Sofia, Bulgaria
             }

   \date{Received May 5, 2026; accepted June 23, 2026}

  \abstract
    {The HD\,202206 system, which features two substellar companions with a 5:1 period ratio around a solar-type star, offers a rare opportunity to study high-order mean-motion resonances and provides new insights into the formation and evolution of substellar companions in extrasolar planetary systems.}
   {We revisited the HD\,202206 system, aiming to conduct a more comprehensive analysis of orbital inclinations, companion masses, resonance dynamics, and potential formation mechanisms.}
   {Our analysis of the astrometric jitter around the best fit in the \textit{Gaia} DR3 catalog and the proper motion anomalies between \textsc{Hipparcos} and \textit{Gaia} places robust constraints on the orbital inclinations. We performed a new dynamical fit to all available radial velocity data from the CORALIE and HARPS spectrographs. We assessed the dynamical configuration and long-term stability of the system using samples generated by nested sampling.}
   {Our analysis shows that the orbital inclinations are strongly constrained to about $51^\circ$. Consequently, the best-fit dynamical solution, assuming a coplanar and inclined ($i = 51^\circ$) configuration, yields an inner brown dwarf of $21.56 M_\mathrm{J}$ and an outer giant planet of $3.12 M_\mathrm{J}$. The corresponding orbital periods are $256.26$ days and $1298.87$ days, with eccentricities of $0.426$ and $0.180$. An investigation of the five relevant resonance angles shows that only one is librating with a large amplitude. Our stability analysis confirms that the system is dynamically stable.}
   {Our results provide new constraints on the masses and inclination of the system, challenging earlier claims of a face-on configuration and shedding light on the formation and evolution of substellar companions in extrasolar planetary systems.}

   \maketitle

\section{Introduction}

    Mean-motion resonances (MMRs) in extrasolar multi-planet systems offer key insights into their formation and dynamical history. In particular, they provide important constraints on the migration and eccentricity damping induced by planet--disk interactions (\citealt{LeePeale2002,Kley2004}). 
    Most observed resonant pairs are found in first-order resonances. Well-established examples include GJ\,876 (\citealt{Marcy2001}), HD\,82943 (\citealt{Mayor2004}), HD\,73526 (\citealt{Tinney2006}), $\eta$\,Cet (\citealt{trifonov_2014}), and HD\,27894 (\citealt{Trifonov2017}) in the 2:1 MMR, as well as HD\,45364 in the 3:2 MMR (\citealt{Correia2009}) and 7 CMa in the 4:3 MMR \citep{Luque2019}.

    In contrast, high-order resonances are comparatively rare. Confirmed examples include the 3:1 MMR in HD\,60532 \citep{Desort2008,3/1}, the 5:1 MMR in HD\,202206 \citep{Correia2005,Couetdic2010}, and the 6:1 MMR in $\nu$ Ophiuchi \citep{sao2012,Quirrenbach2019}. Several additional systems have been reported with period ratios close to high-order commensurabilities, such as 4:1 in HD\,108874 (\citealt{4/1}), 7:5 in HD\,41248 (\citealt{7/5}), and 9:7 in the Kepler-29 system (\citealt{9/7}). The scarcity of high-order MMRs likely reflects the increasingly restrictive dynamical conditions required for resonant capture and long-term stability as the resonance order increases, making such configurations more difficult to establish and preserve during planetary migration. These cases challenge standard migration scenarios and help constrain the diversity of disk conditions and dynamical histories (\citealt{Papaloizou}).

    The inner companion of HD\,202206 has a minimum mass $\approx 17~M_\text{J}$ (where $M_\text{J}$ is the mass of Jupiter) and is thus firmly in the brown dwarf regime; the outer companion is a giant planet \citep{Correia2005,Couetdic2010}. A resonant system that contains both a brown dwarf and a giant planet provides valuable insight into the limits of planet and brown dwarf formation, and into the possibility of a continuous transition between the two populations. Brown dwarfs can form either through fragmentation like low-mass stars or within disks like massive planets, and their coexistence with resonant giant planets offers an important test of these different scenarios \citep{Mordasini2009,Ma2014}.

    Previous studies of HD\,202206 have established the dynamical structure around the 5:1 commensurability (\citealt{Correia2005}; \citealt{Couetdic2010}). Combining \textit{Hubble} Space Telescope Fine Guidance Sensor astrometry with radial velocities (RVs), \cite{Benedict2017} determined orbital inclinations of $i_b = 10.9^\circ \pm 0.8^\circ$ and $i_c = 7.7^\circ \pm 1.1^\circ$ for the inner and outer companions, respectively. This indicates that the system is nearly face on and approximately coplanar, and implies a stellar-mass inner companion and a brown dwarf outer companion.

    In this work, we reexamined the inclination and show that a nearly face-on solution is not supported. We refined constraints on the true architecture of HD\,202206 and confirmed the substellar nature of the inner companion. The CORALIE and High Accuracy Radial velocity Planet Searcher (HARPS) RV data and the parameters of the host star are provided in Sects.~\ref{sec2} and \ref{sec3}, respectively. In Sect.~\ref{sec4} we present the Keplerian and coplanar dynamical fits derived from the RV data. Then the inclination constraints are detailed, with the inclination of the system constrained to near $51^\circ$, and we present the best coplanar $i = 51^\circ$ dynamical fit. The orbital evolution and long-term dynamical stability of the system are examined in Sect.~\ref{sec6}. In Sect.~\ref{sec7} we discuss why only one resonant angle is found to librate, as well as the formation and migration scenario for this system. Finally, we summarize our results in Sect.~\ref{sec8}.

\section{Radial velocity data}
\label{sec2}
     The HD\,202206 system was discovered in the CORALIE survey \citep{Correia2005,Couetdic2010}. The CORALIE \'echelle spectrograph is mounted on the Swiss 1.2-meter \textit{Leonhard Euler} Telescope at La Silla Observatory of the European Southern Observatory (ESO).\ It has an average measurement precision of $\sim 5$\,m\,s$^{-1}$ \citep{Coralie}. A total of 92 RV measurements were obtained between 1999 and 2006, with a median internal measurement error of $5.8$\,m\,s$^{-1}$. 

    We also found 86 RV measurements from the HARPS spectrograph (obtained between 2004 and 2021) in the spectra available in the public ESO archive\footnote{\url{https://archive.eso.org/wdb/wdb/adp/phase3_main/form}} and extracted via the HARPS-RVBank \citep{Trifon2020}. HARPS,  in operation at the ESO La Silla 3.6m telescope, is a second-generation RV spectrograph that can attain a higher RV precision of $\sim 1$\,m\,s$^{-1}$. These measurements have much smaller internal errors (with a median of $1.0$\,m\,s$^{-1}$) and significantly expand the time span of observations. 

Although the  CORALIE RV data are less precise, they increase the time span of available observations by more than 4 years. As a result, our analysis of this system was conducted with the combined dataset of the CORALIE and HARPS data. HARPS underwent a fiber exchange in 2015, which introduced a small but significant RV offset between the two epochs. Hence, its observational record is divided into the HARPS-pre and HARPS-post datasets. The HARPS-pre dataset initially consisted of 64 individual RV measurements. Multiple observations obtained on the same night are not statistically independent and may be affected by short-timescale stellar variability and instrumental systematics. To reduce the impact of these effects and to avoid over-weighting densely sampled nights, we computed nightly averaged RVs. This binning reduced the HARPS-pre dataset to 59 independent data points. Similarly, the HARPS-post dataset was reduced from 22 individual measurements to 20 nightly averaged data points. In total, the combined dataset consists of 171 RVs.

\section{Stellar parameters}
\label{sec3}

\begin{figure}[t]
\centering
\includegraphics[width=0.85\hsize]{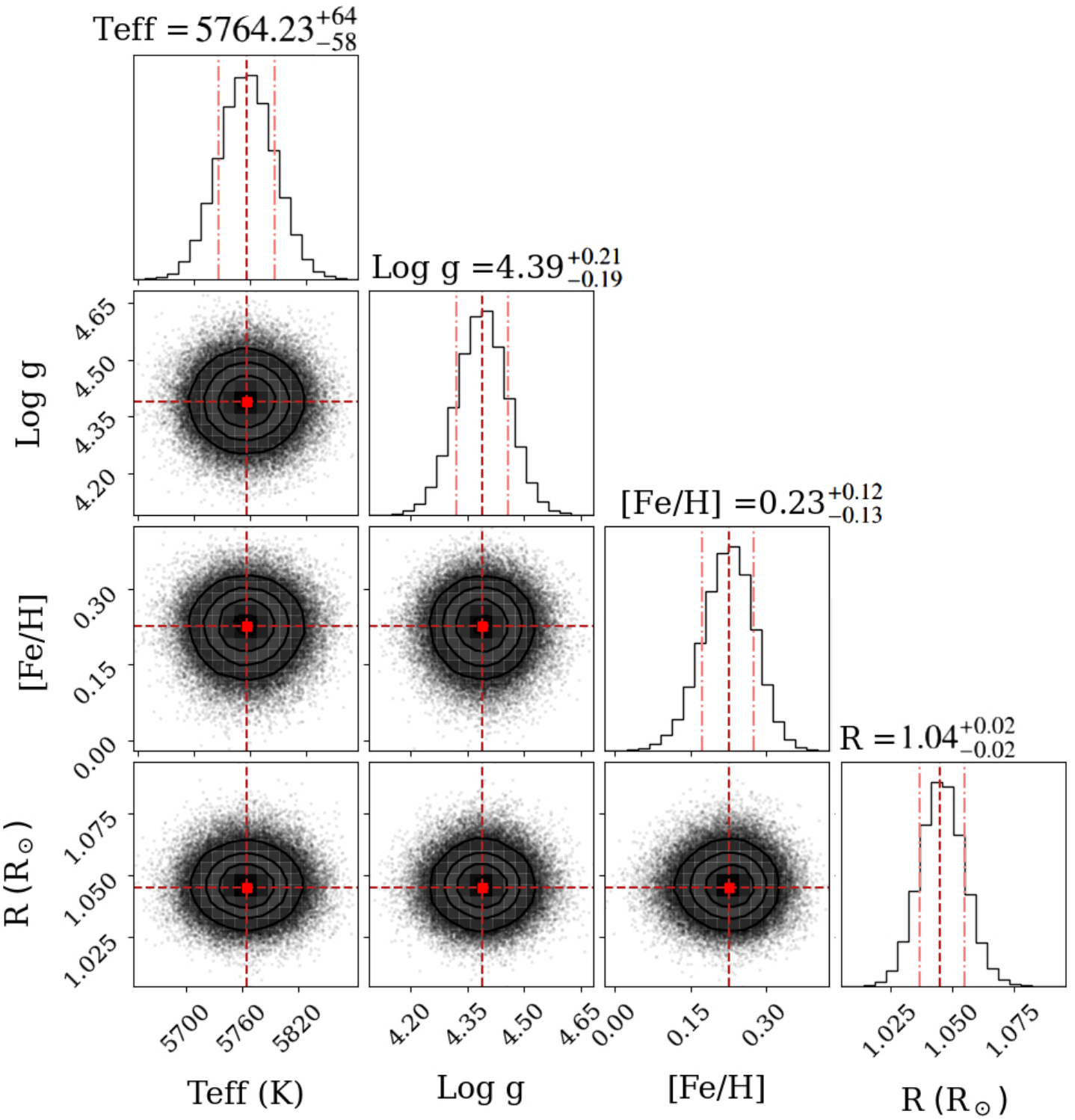}
    \caption{Posterior probability distribution of the effective temperature ($T_{\text{eff}}$), surface gravity ($\log g$), metallicity ([Fe/H]), and stellar radius ($R$) of the star HD\,202206. The dashed red lines and squares indicate the median values and the $1\sigma$ confidence intervals.}
        \label{StarCORNER}
\end{figure} 

HD\,202206 (HIP 104903) is a photometrically stable G2V main-sequence star. We derived the stellar properties of HD\,202206 by modeling its spectral energy distribution (SED) using the code ARIADNE\footnote{\url{https://github.com/jvines/astroARIADNE}} \citep{ARIADNE}. This tool adopts a Bayesian framework and incorporates multiple SED model grids in order to reduce biases associated with individual model choices. It also estimates stellar mass and age via interpolation of MESA Isochrones and Stellar Tracks \citep[MIST;][]{Dotter2016, Choi2016}. The SED fitting was performed using photometric data from the following bands: GALEX far-UV; Str\"omgren v, b, and y; Johnson B and V; Tycho B and V; \textit{Gaia} Data Release 2 G, BP, and RP; TESS (Transiting Exoplanet Survey Satellite); 2MASS J, H, and K; and WISE W1 and W2. These measurements were fitted using atmosphere models from PHOENIX v2 \citep{Husser2013}, BT-Settl, BT-NextGegn, BT-Cond \citep{Hauschildt1999, Allard2012}, \cite{Castelli&Kurucz2003}, and \cite{Kurucz1993}.

We adopted a Gaussian prior on the distance based on the \textit{Gaia} Early Data Release 3 estimate from \cite{Bailer-Jones2021}. Interstellar extinction was assigned a uniform prior between zero and the maximum extinction along the line of sight, computed using the Galactic dust maps of \citet{Schlafly&Finkbeiner2011} in the Python package \textit{dustmaps}\footnote{\url{https://github.com/gregreen/dustmaps}} \citep{Green2018}. For the initial stellar parameters ($T_{\text{eff}}$, $\log{g}$, and [Fe/H]), we imposed Gaussian priors centered on the values reported by \citet{Perdelwitz2024} in the HARPS-RVBank. These quantities were derived from templates constructed by co-adding HARPS spectra (see \citealp{Perdelwitz2024}, Sect.~2.1).

A complete overview of the derived parameters is provided in Table~\ref{stellar param}, and the marginalized posterior distributions of some of the parameters are shown in Fig.~\ref{StarCORNER}. The resulting stellar parameters are consistent with recent literature values \citep[e.g.,][]{Stassun2017,Stassun2018}. In particular, we find that the stellar mass is $1.06_{-0.06}^{+0.07}\,M_\odot$, compared to $1.15\,M_\odot$ in \cite{Correia2005}, $1.044\,M_\odot$ in \cite{Couetdic2010}, $1.27\pm 0.12\,M_\odot$ in \cite{Stassun2017}, and $1.03_{-0.11}^{+0.14}\,M_\odot$ in the TESS Input Catalog \citep{Stassun2018}.

\begin{table}[t]
\centering
\caption{Stellar parameters of HD\,202206.}
\label{stellar param}
\renewcommand{\arraystretch}{1.35}
\setlength{\tabcolsep}{6pt} 
\begin{tabular}{lcr}
\hline\hline
Parameter                   & Value                   & Reference    \\ \hline
Parallax {[}mas{]}                  & $21.9392 \pm 0.0275$            & Gaia DR3$^a$     \\
$\mu_\alpha \cos \delta$ {[}mas yr$^{-1}${]}  & $-39.173\pm0.027$       & Gaia DR3     \\
$\mu_\delta$ {[}mas yr$^{-1}${]} & $-120.068\pm0.022$      & Gaia DR3     \\
$V$ {[}mag{]}     & $8.07\pm0.01$           & Tycho-2$^b$      \\
$B$ {[}mag{]}   &   $8.79 \pm 0.02$ & Tycho-2   \\ \hline
Stellar type                & G2V                     & This work    \\
$T_\mathrm{eff}$ [K]        & $5764.23^{+64}_{-58}$   & This work    \\
$\log g$ [cgs]              & $4.39^{+0.21}_{-0.19}$  & This work    \\
{[}Fe/H{]}                  & $0.23^{+0.12}_{-0.13}$  & This work    \\
Radius [$R_\odot$]          & $1.04^{+0.02}_{-0.02}$  & This work    \\
Luminosity [$L_\odot$]      & $1.09^{+0.06}_{-0.06}$  & This work    \\
Age [Gyr]                   & $3.94^{+1.28}_{-3.91}$  & This work    \\
Mass [$M_\odot$]            & $1.06^{+0.07}_{-0.06}$  & This work    \\
Distance {[}pc{]}           & $45.62^{+0.10}_{-0.06}$ & This work    \\\hline
\end{tabular}
\tablefoot{$^a$\citet{Gaia2023}; $^b$\citet{Tycho-2}.}
\end{table}

\section{Orbital solutions from radial velocity and astrometric analysis}
\label{sec4}

\subsection{Keplerian fit}

    To model the orbital configuration of the system, we first assumed that the companions are on non-interacting Keplerian orbits. We adopt JD = 2451402.808 (the first CORALIE RV measurement) as the reference epoch throughout this paper. Within the {\sc Exo-Striker}\footnote{\url{https://github.com/3fon3fonov/exostriker}} exoplanet toolbox \citep{Trifonov2019}, we used the Nelder–Mead (Simplex) algorithm (\citealt{Nelder1965}) to optimize the likelihood function ($\ln \mathcal{L}$) coupled with a two-planet Keplerian model. The parameters that were derived through the fitting are the RV semi-amplitude ($K$), orbital period ($P$), eccentricity ($e$), argument of periastron ($\omega$), mean anomaly ($M$), and the RV data offset (RV$_{\text{off}}$) and jitter (RV$_{\text{jit}}$) for each dataset. All the orbital parameters are given in the Jacobi coordinate system \citep{LeePeale2003}. The best Keplerian fit to the combined dataset has $\ln \mathcal{L} = -598.02$ and is consistent with two companions of minimum masses $m_b\sin i = 16.70\,M_J$ and $m_c\sin i = 2.43\,M_{J}$, orbital periods $P_b = 256.36$ days and $P_c = 1291.02$ days, and eccentricities $e_b = 0.426$ and $e_c =0.130$ (Table \ref{tab:combined_fits}).

\begin{table*}[ht]
    \centering
    \caption{Best Keplerian and coplanar dynamical fits to the HARPS and CORALIE RV data of HD\,202206.}
    \label{tab:combined_fits}
    \renewcommand{\arraystretch}{1.3}
    \begin{tabular*}{\textwidth}{@{\extracolsep{\fill}} l c c c c c c @{}}
        \hline\hline
        \raisebox{-1.5ex}[0pt]{Orb. Param.} & \multicolumn{2}{c}{Keplerian fit ($i = 90^\circ$)} & \multicolumn{2}{c}{Dynamical fit ($i = 90^\circ$)} & \multicolumn{2}{c}{Dynamical fit ($i = 51^\circ$)} \\
        \cline{2-3} \cline{4-5} \cline{6-7}
        & Inner & Outer & Inner & Outer & Inner & Outer \\
        \hline
        $K$ (m\,s$^{-1}$) & 562.37 & 43.55 & 562.61 & 43.61 & 562.64 & 43.51 \\
        $P$ (d)         & 256.36 & 1291.02 & 256.28 & 1295.03 & 256.26 & 1298.87 \\
        $e$             & 0.426   & 0.130   & 0.426   & 0.177    & 0.426   & 0.180    \\
        $\omega$ ($^\circ$)     & 161.80 & 86.16  & 161.65 & 81.27   & 161.66 & 82.73   \\
        $M$ ($^\circ$)         & 354.52 & 80.37   & 354.38 & 84.25   & 354.33 & 83.76   \\
        $a$ (AU)        & 0.81   & 2.38    & 0.81   & 2.38    & 0.81   & 2.39   \\
        $m$ $(M_\text{J})$   & 16.70  & 2.43    & 16.70  & 2.42    & 21.56  & 3.12    \\
        RV$_{\text{off. CORALIE}}$ (m\,s$^{-1}$) & \multicolumn{2}{c}{726.39} & \multicolumn{2}{c}{726.82} & \multicolumn{2}{c}{726.71} \\
        RV$_{\text{off. HAPRS-pre}}$ (m\,s$^{-1}$) & \multicolumn{2}{c}{$-15.65$} & \multicolumn{2}{c}{$-14.69$} & \multicolumn{2}{c}{$-14.59$} \\
        RV$_{\text{off. HAPRS-post}}$ (m\,s$^{-1}$) & \multicolumn{2}{c}{$-2.39$} & \multicolumn{2}{c}{$-3.24$} & \multicolumn{2}{c}{$-2.76$} \\  
        RV$_{\text{jit. CORALIE}}$ (m\,s$^{-1}$) & \multicolumn{2}{c}{6.45} & \multicolumn{2}{c}{6.62} & \multicolumn{2}{c}{6.53} \\
        RV$_{\text{jit. HAPRS-pre}}$ (m\,s$^{-1}$) & \multicolumn{2}{c}{7.21} & \multicolumn{2}{c}{7.26} & \multicolumn{2}{c}{7.31} \\
        RV$_{\text{jit. HAPRS-post}}$ (m\,s$^{-1}$) & \multicolumn{2}{c}{6.09} & \multicolumn{2}{c}{6.45} & \multicolumn{2}{c}{6.58} \\ 
        rms (m\,s$^{-1}$) & \multicolumn{2}{c}{8.20} & \multicolumn{2}{c}{7.29} & \multicolumn{2}{c}{7.99} \\        
        $\ln\mathcal{L}$ & \multicolumn{2}{c}{$-598.02$} & \multicolumn{2}{c}{$-595.31$} & \multicolumn{2}{c}{$-594.73$} \\
        \hline
    \end{tabular*}
\end{table*}

\subsection{Coplanar dynamical fits}    
    For an investigation of the mutual gravitational interactions between massive companions, a description based solely on an independent Keplerian fit is inadequate. Such an approach neglects the planet–planet perturbations that can significantly modify the orbital evolution. A self-consistent $N$-body treatment is therefore required to properly characterize the system architecture and assess its long-term dynamical behavior. Hence, we used the best Keplerian fit as an initial guess and performed a maximum likelihood fit coupled with a self-consistent dynamical model. We first assumed a coplanar and edge-on configuration of the system ($i_b = i_c = 90^\circ$, $\Delta\Omega = 0^\circ$). The best-fit parameters at the reference epoch are listed in Table~\ref{tab:combined_fits}. This fit has $\ln\mathcal{L} = -595.31$, which is slightly better than the Keplerian fit, and the resulting parameters are similar to the Keplerian fit, except for the slightly longer period ($P_c = 1295.03$ days) and large eccentricity ($e_c = 0.177$) of the outer companion.

    Next, we aimed to constrain the inclination of the HD\,202206 system by examining the relationship between inclination and the quality of the RV fit, using specifically the log likelihood ($\ln \mathcal{L}$) and the Bayesian information criterion (BIC). These metrics provide a quantitative evaluation of the quality of the dynamical fits to the observed RV data.

\begin{figure}[t]
\centering
\includegraphics[width=0.95\hsize]{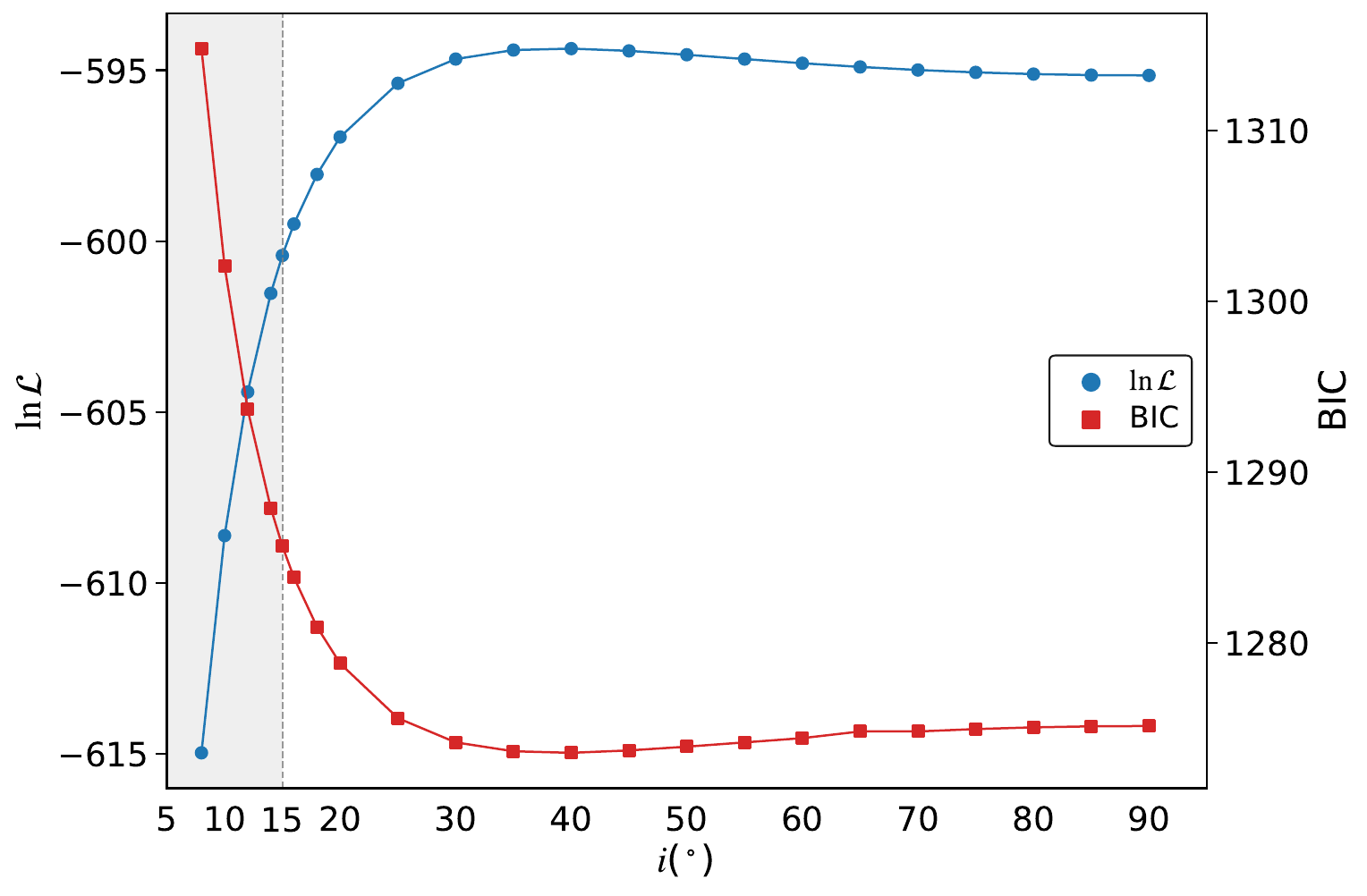}
    \caption{Resulting ln $\mathcal{L}$ and BIC for coplanar dynamical fits as a function of the inclination ($i$) for the combined HARPS and CORALIE dataset. The left-hand axis is for $\ln \mathcal{L}$ shown as blue circles and the right-hand axis for BIC shown in red squares. The shaded gray region illustrates the inclination that should be ruled out.}
        \label{BIC}
\end{figure}

    We systematically varied the system inclination ($i$) from $90^\circ$ down to $8^\circ$ and recorded the corresponding fitting statistics, $\ln \mathcal{L}$ and BIC, which we present in Fig.~\ref{BIC}. The figure shows both the $\ln \mathcal{L}$ and BIC. As we can see, varying the inclination from $90^\circ$ to $30^\circ$ results in minimal changes to the quality of the RV fit. For inclinations below $25^\circ$, both $\ln \mathcal{L}$ and BIC vary rapidly: $\ln \mathcal{L}$ decreases and BIC increases, indicating a strong degradation of the model at low inclinations. The global maximum of $\ln \mathcal{L}$ and the global minimum of BIC for the combined CORALIE and HARPS data occur at $i = 40^\circ$, and we adopted those values as ln $\mathcal{L}_{\text{max}}$ and BIC$_{\text{min}}$ when computing differential metrics.
    
    To identify unfavorable inclination in the fitting, we defined the following metrics: $\Delta\ln \mathcal{L} = |\ln \mathcal{L}| - |\ln \mathcal{L}_{\text{max}}|$ and $\Delta\text{BIC} = \text{BIC} - \text{BIC}_\text{{min}}$.  The shaded gray region highlights inclinations that simultaneously breach the $\Delta\ln \mathcal{L} > 7$ and $\Delta\text{BIC} > 10$ thresholds \citep{kass1995bayes, Trifon2022}. The above tests reveal that  an inclination $\leq$ 15° is strongly disfavored by the RV fits, contradicting the low‑inclination solution reported by \citet{Benedict2017}.

\subsection{Astrometric analysis}

    To further constrain the inclination and the three-dimensional geometry of the 
    HD\,202206 system, we made use of the \textit{Gaia} Data Release 3 (DR3) astrometry together with the
    RV posteriors from a nested sampling run of the Keplerian fit.
    We modeled the astrometric orbits of the two companions under a coplanar assumption 
    with the same inclinations and longitudes of the ascending nodes. 
    We used three constraints to fit the orbits: the astrometric jitter around the best 
    fit in the \textit{Gaia} DR3 catalog (assuming this is exclusively due to the companions)
    as well as the two proper motion anomalies in right ascension and declination.
    The latter are derived from the differences between the catalog proper motions from
    Hipparcos \citep{vanleeuwen07} and \textit{Gaia} DR3, respectively, and the long-term
    proper motion computed from the positional
    difference between \textit{Hipparcos} and \textit{Gaia} DR3, as given by \citet{GAIADR3}.
    Because the \textit{Gaia} DR3 astrometric jitter includes epoch-astrometry outliers that 
    have not yet been removed, it should be regarded as an upper limit on the true companion-
    induced signal; consequently, the inclination derived from the astrometric jitter should 
    be interpreted as a lower limit; the true absolute inclination may be higher than the 
    nominal value inferred here.
    
    Briefly, we fitted the astrometric orbits by drawing from the spectroscopic posteriors
    1000 times. For each
    of those 1000 combinations of spectroscopic parameters, we computed the log likelihood
    by combining the three different constraints from the astrometry and performed a nested
    sampling run using \textsc{dynesty} \citep{dynesty}, in order 
    to obtain the best inclination and ascending node. Finally, we obtained 
    a single posterior that takes both the errors in the spectroscopic
    parameters and in the astrometric constraints into account. In order to simulate the astrometric orbit and compare the resulting astrometric excess jitter with that reported 
    in \textit{Gaia} DR3, we obtained the time and scan direction of the individual 
    \textit{Gaia} abscissa measurements from the \textit{Gaia} Observation Forecast 
    Tool GOST\footnote{https://gaia.esac.esa.int/gost/},
    and we estimated the accuracy of each
    \textit{Gaia} measurement from the final catalog accuracy and the number of individual
    measurements that went into it. We also fitted positions, proper motions, and
    parallax terms to our simulated abscissae so that the simulated astrometric excess
    scatter does not contain terms that would have been subsumed into those parameters.

\begin{figure}
\centering
\includegraphics[width=0.8\hsize]{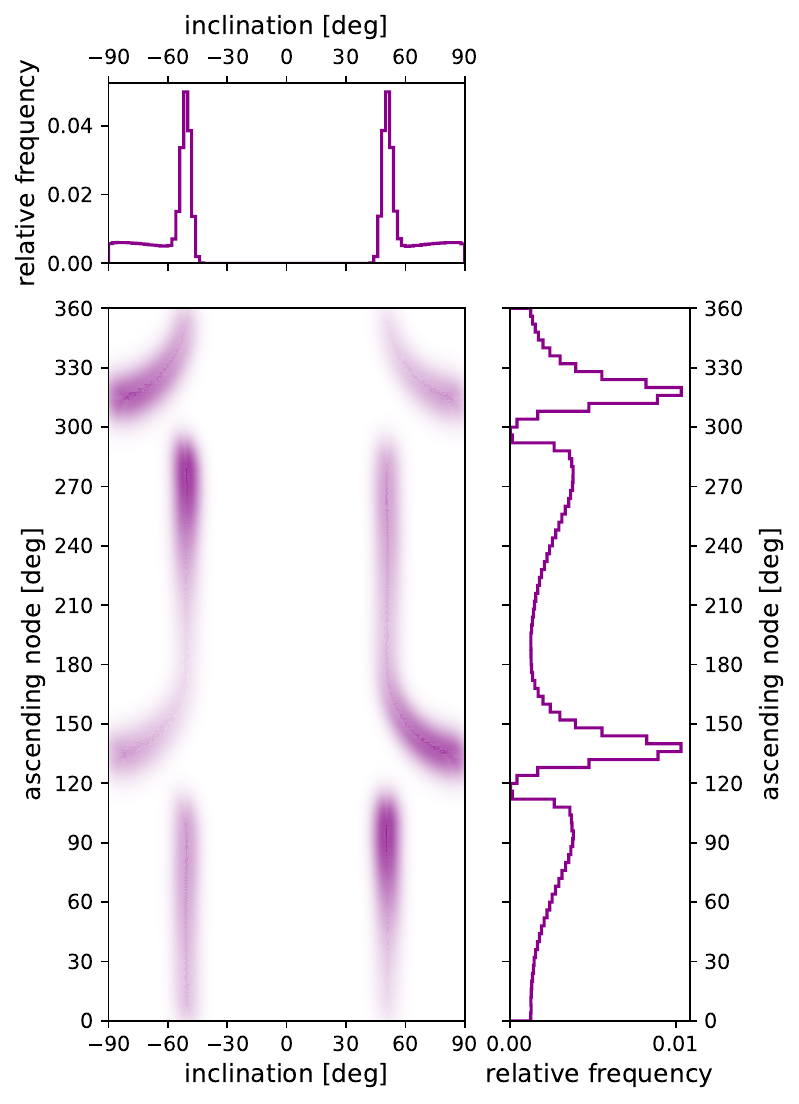}
\caption{Posterior probability distributions for the orbital inclination ($i$) and the longitude of the ascending node ($\Omega$) derived from the nested sampling analysis using astrometric constraints. \textit{Center}: Joint density distribution. \textit{Top and right}: Marginalized one-dimensional probability density functions. The inclination peaks at approximately $\pm 50.7^\circ$, and the longitude of the ascending node presents corresponding peaks at $137.5^\circ$ and $317.5^\circ$.}
\label{inclinations posteriors}
\end{figure}

    Figure~\ref{inclinations posteriors} shows the posterior distribution of the
    inclination and ascending nodes of the astrometric analysis, and Table~\ref{ascendingnode}
    provides the corresponding values of the resulting inclination and ascending nodes.
    The orbital inclination derived from our simulations converges consistently to values
    near $51^\circ$. 
    The posterior distributions exhibit a sharp and asymmetric peak, with a best-fit value
    of $50.7^{+25.8}_{-1.4}$ degrees and a long shallow tail toward higher inclinations.
    The narrow widths of the posterior peaks indicate that
    the inclination is well constrained once the spectroscopic and astrometric data are
    combined. 
    
\begin{figure*}[ht]
\centering
\includegraphics[width=0.85\hsize]{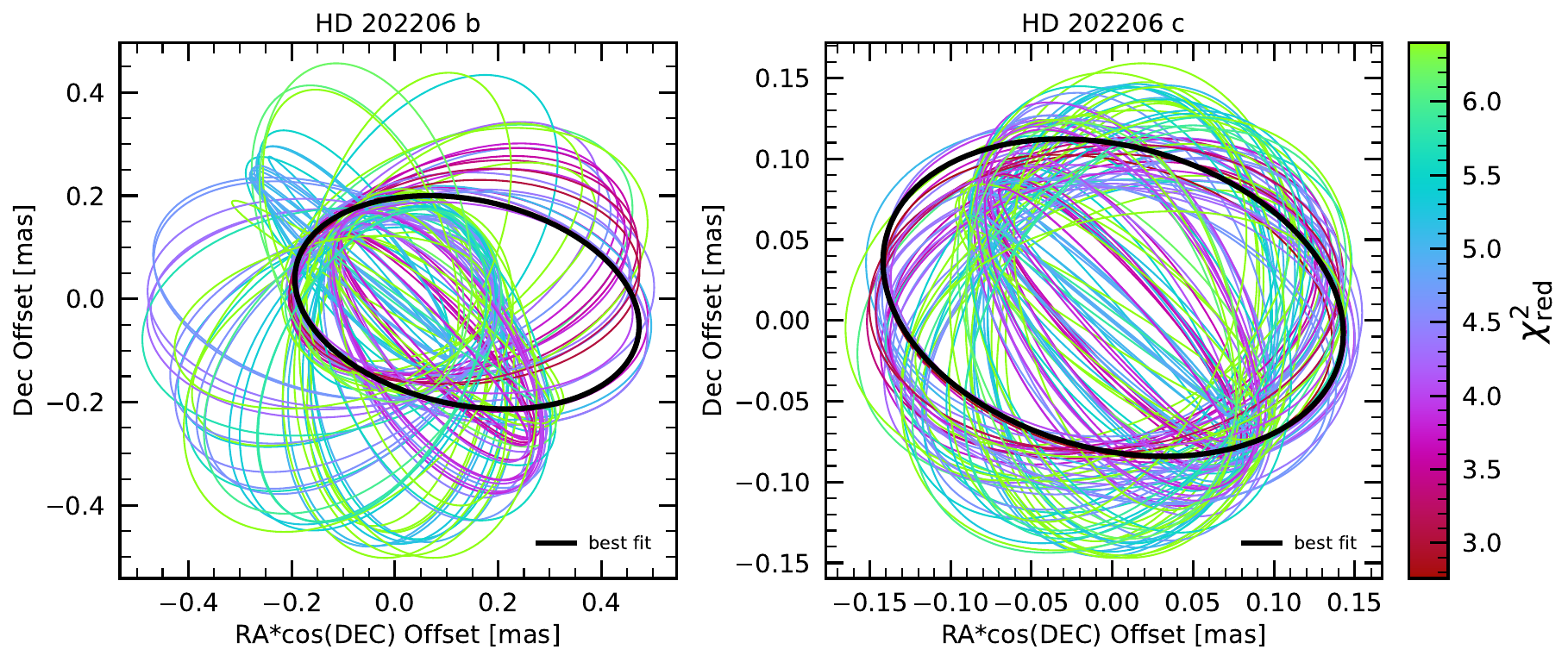}
\caption{Astrometric orbital simulations for HD\,202206 b (\textit{left}) and HD\,202206 c (\textit{right}) constrained by astrometric data. The simulations utilize the proper motion anomalies in right ascension and declination, defined as the difference between the \textsc{Hipparcos} and \textit{Gaia} proper motion vectors, as reported by \citet{GAIADR3}. The curves represent a sample of orbital solutions drawn from the posterior distribution, color-coded by the reduced chi-squared statistic ($\chi^2_{\mathrm{red}}$) of the fit to the astrometric constraints. The solid black ellipses indicate the best-fitting orbital solutions corresponding to the minimum $\chi^2_{\mathrm{red}}$.}
\label{astrometric orbit}
\end{figure*}

    Together with the inclination, the longitude of the ascending node is also restricted to
    two groups of solutions around $138^\circ$ and $318^\circ$. 
    Figure~\ref{astrometric orbit} illustrates the fitted astrometric 
    orbits of the two companions in the HD\,202206 system.

\begin{table}[t]
\caption{Inclination and ascending node of the HD\,202206 system from astrometric analysis.}
\label{ascendingnode}
\centering
\renewcommand{\arraystretch}{1.5}
\begin{tabular*}{\linewidth}{@{\hspace{0.5em}} l @{\extracolsep{\fill}} c @{\hspace{0.5em}}}
\hline\hline
Parameter & Value$^a$ \\
\hline
Inclination (positive)        & $50.7^{+25.8}_{-1.4}$ \\
Inclination (negative)        & $-50.7^{+1.4}_{-25.8}$ \\
\hline
Ascending node (small) & $137.5^{+8.6}_{-84.9}$ \\
Ascending node (large) & $317.5^{+8.6}_{-85.0}$ \\
\hline
\end{tabular*}
\tablefoot{$^a$Posterior modes with 1$\sigma$ confidence intervals.}
\end{table}

\subsection{Coplanar dynamical fit with $i = 51^\circ$}    
\begin{figure*}
\centering
\includegraphics[width=0.75\hsize]{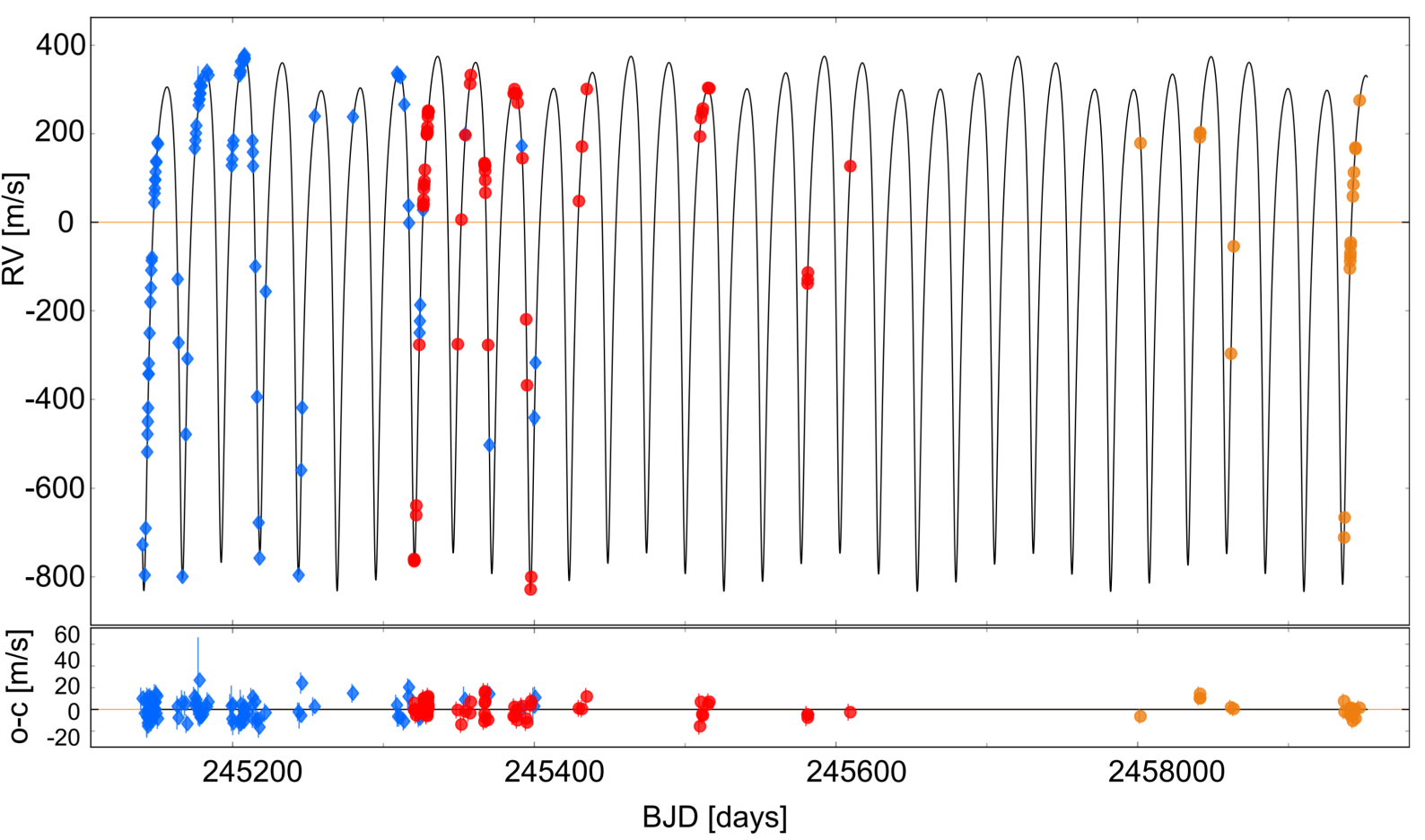}
    \caption{\textit{Top}: Best coplanar $i = 51^\circ$ dynamical fit to the CORALIE (blue), HARPS-pre (red), and HARPS-post (orange) RV data of HD\,202206. \textit{Bottom:} Residuals around the best fit.}
        \label{RVdata}
\end{figure*}

    Since our astrometric analysis showed that the orbital inclination is strongly constrained to about $51^{\circ}$, we derived the best coplanar inclined fit by fixing the inclination to $51^{\circ}$ and redoing the dynamical fit to the CORALIE and HARPS RV data. Figure~\ref{RVdata} shows the best coplanar $i = 51^\circ$ dynamical fit, and the parameters derived for this fit are listed in Table~\ref{tab:combined_fits}. Our best-fit dynamical solution yields actual masses of $21.56\,M_\text{J}$ and $3.12\,M_\text{J}$ for the inner and outer companions, respectively. The corresponding orbital periods are 256.26 days and 1298.87 days, with eccentricities of 0.426 and 0.180, respectively, at the reference epoch.

\section{Orbital evolution}
\label{sec6}
\subsection{Evolution of the best $i=51^\circ$ fit}
    In this section we begin by presenting the evolution of key orbital parameters for the HD\,202206 system. The integrator we used for our $N$-body simulation is the Wisdom-Holman symplectic integrator \citep{Wisdom1991} implemented in the {\sc Exo-Striker}. Starting from the initial observational epoch JD = 2451402.808 (the first CORALIE data point), we performed the orbital integration for 1 million years using a fixed time step of 2 days, which gives roughly 128 integration steps per orbit of the inner companion.

    Given that the orbital period ratio of the companions is approximately 5:1, we further monitored both the evolution of the secular apsidal angle $(\Delta\varpi = \varpi_b - \varpi_c)$ and the evolution of all five resonance angles, defined as follows:
    \begin{align}
    \theta_1 &= \lambda_b - 5\lambda_c + 4\varpi_b, \label{eq:theta1} \\
    \theta_2 &= \lambda_b - 5\lambda_c + 3\varpi_b + \varpi_c, \label{eq:theta2} \\
    \theta_3 &= \lambda_b - 5\lambda_c + 2\varpi_b + 2\varpi_c, \label{eq:theta3} \\
    \theta_4 &= \lambda_b - 5\lambda_c +  \varpi_b + 3\varpi_c, \label{eq:theta4} \\
    \theta_5 &= \lambda_b - 5\lambda_c + 4\varpi_c, \label{eq:theta5} 
    \end{align}
    where $\varpi_{b,c} = \Omega_{b,c} + \omega_{b,c}$ is the longitude of periastron and $\lambda_{b,c} = M_{b,c} + \varpi_{b,c}$ is the mean longitude, with $\Omega_{b,c}$, $\omega_{b,c}$, and $M_{b,c}$ being the longitude of the ascending node, the argument of periastron, and the mean anomaly, respectively. Note that $\Omega_b = \Omega_c = 0^\circ$ for our coplanar fits.

\begin{figure*}[ht]
\centering
\includegraphics[width=0.9\hsize]{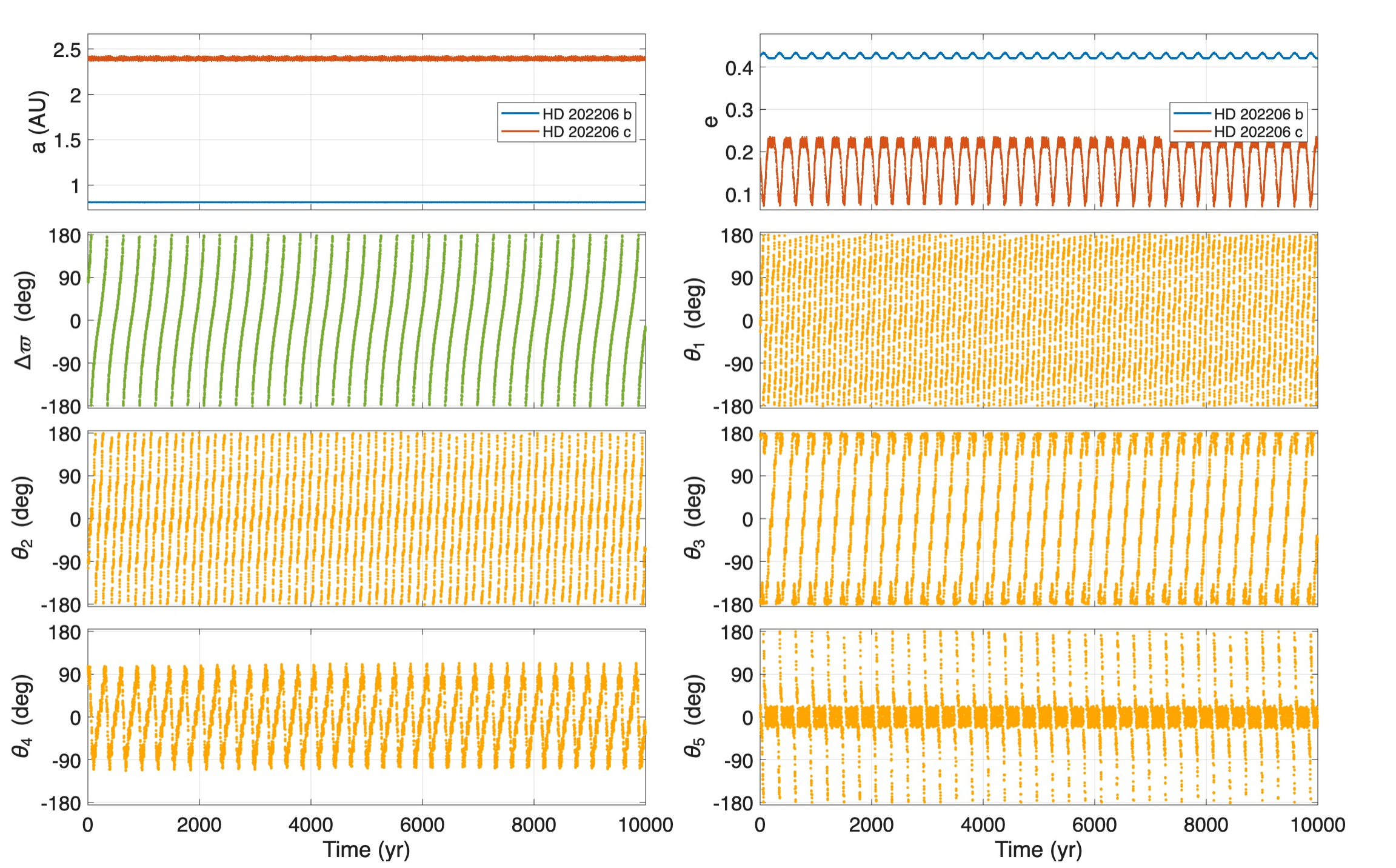}
    \caption{Orbital evolution of the best coplanar inclined ($i = 51^\circ$) dynamical fit, shown for 10 kyr from the $N$-body simulation. The panels display the semimajor axes $a_b$ (blue) and $a_c$ (red), the eccentricities $e_b$ (blue) and $e_c$ (red), the secular apsidal difference, $\Delta\varpi$ (green), and the resonant angles $\theta_1$, $\theta_2$, $\theta_3$, $\theta_4$, and $\theta_5$ (orange).}
        \label{Evolution}
\end{figure*}

     Figure~\ref{Evolution} shows a 10 kyr segment drawn from the 1 Myr integration of the best coplanar $i = 51^\circ$ fit. Over the entire 10 kyr window, the semimajor axes are effectively constant, showing only small, regular variations. The inner companion remains around $a_b$ $\approx$ 0.80 AU, and the outer companion remains near $a_c$ $\approx$ 2.40 AU. The eccentricities display coherent, periodic variations on a timescale of $\approx 290$\,yr. The eccentricity of the inner companion exhibits small-amplitude oscillations with $e_b \approx 0.41$–-$0.43$ (mean $\approx$ 0.42). The outer companion shows larger cyclic variations with $e_c$ $\approx$ 0.06–-0.22 (mean $\approx$ 0.15–0.16) over each cycle. These eccentricity oscillations are regular and repeat across the entire 1 Myr interval. The apsidal difference ($\Delta\varpi$) circulates through the full ±180° range on the same timescale of $\approx 290$\,yr (green), indicating that the orbits are not locked in an aligned apsidal configuration. Of the five resonant angles, only $\theta_4$ clearly librates around 0°, with a semi‑amplitude of approximately $\pm 100$–-$120^\circ$. The other angles, $\theta_1$, $\theta_2$, $\theta_3$, and $\theta_5$, circulate through the full $\pm 180^\circ$ range. The libration period seen in $\theta_4$ is consistent with the dominant period of the eccentricity oscillations and the circulation of $\Delta\varpi$.

    \begin{table*}[ht]
    \centering
    \caption{Posterior estimates (median and $1\sigma$ region) for coplanar edge-on and inclined ($i$ = 51°) dynamical fits.}
    \label{orbital param}
    \renewcommand{\arraystretch}{1.5} 
    \begin{tabular*}{\linewidth}{@{\hspace{0.5em}} l @{\extracolsep{\fill}} c c c c c c@{\hspace{0.5em}}}
    \hline\hline 
    & \multicolumn{2}{c}{N-body($i = 51^\circ$)}   &  \multicolumn{2}{c}{N-body($i = 90^\circ$)} &   \multicolumn{2}{c}{Adopted Priors}\\
    \hline
    Orb. Param. & Inner & Outer & Inner & Outer & Inner & Outer \\
    \hline
    $K$ (m s$^{-1}$) & $562.58^{+1.35}_{-1.36}$  & $43.04^{+1.17}_{-1.16}$   & $560.87^{+3.04}_{-0.31}$  & $44.02^{+0.05}_{-2.20}$    & $\mathcal{U}$(500, 600)  & $\mathcal U$(40,50)  \\
    $P$ (d)         & $256.26^{+0.03}_{-0.03}$  & $1299.22^{+3.21}_{-2.76}$  & $256.32^{+0.01}_{-0.08}$  & $1294.31^{+6.59}_{-0.94}$  & $\mathcal U$(200,300)  & $\mathcal U$(1200,1400)  \\
    $e$             & $0.426^{+0.002}_{-0.002}$  & $0.171^{+0.021}_{-0.025}$       &  $0.427^{+0.001}_{-0.002}$    & $0.149^{+0.039}_{-0.009}$     & $\mathcal U$(0.2,0.6)  & $\mathcal U$(0.01,0.3)  \\
    $\omega$(°)     & $161.64^{+0.29}_{-0.28}$ & $84.73^{+6.53}_{-4.05}$    & $161.82^{+0.11}_{-0.46}$  & $82.86^{+11.20}_{-3.16}$    & $\mathcal U$(0,360)  & $\mathcal U$(0,360)  \\
    $Ma$(°)         & $354.33^{+0.29}_{-0.29}$  & $81.97^{+6.53}_{-6.51}$   & $345.38^{+0.27}_{-0.32}$  & $83.60^{+2.52}_{-10.45}$    & $\mathcal U$(0,360)  & $\mathcal U$(0,360)  \\
    $a$ (AU)        & $0.81^{+0.01}_{-0.02}$    & $2.39^{+0.04}_{-0.05}$        & $0.82^{+0.00}_{-0.02}$  & $2.40^{+0.03}_{-0.06}$    &   \multicolumn{2}{c}{(derived)} \\
    $m$ $(M_\text{J})$   & $21.55^{+0.80}_{-0.81}$   & $3.09^{+0.14}_{-0.14}$   & $16.89^{+0.42}_{-0.83}$ & $2.49^{+0.00}_{-0.20}$    &   \multicolumn{2}{c}{(derived)} \\
    RV$_{\text{off. CORALIE}}$ (m s$^{-1}$) & \multicolumn{2}{c}{$725.94^{+1.92}_{-0.27}$}  & \multicolumn{2}{c}{$726.60^{+1.40}_{-0.83}$}  &   \multicolumn{2}{c}{$\mathcal U$(-1000,1000)}   \\
    RV$_{\text{off. HAPRS-pre}}$ (m s$^{-1}$) & \multicolumn{2}{c}{$-14.21^{+0.66}_{-1.73}$}  & \multicolumn{2}{c}{$-16.56^{+2.81}_{-0.42}$}  &   \multicolumn{2}{c}{$\mathcal U$(-100,100)}   \\
    RV$_{\text{off. HAPRS-post}}$ (m s$^{-1}$) & \multicolumn{2}{c}{$1.81^{+0.71}_{-7.98}$}   & \multicolumn{2}{c}{$-2.38^{+3.70}_{-4.20}$}   &   \multicolumn{2}{c}{$\mathcal U$(-100,100)}   \\  
    RV$_{\text{jit. CORALIE}}$ (m s$^{-1}$) & \multicolumn{2}{c}{$6.80^{+0.59}_{-1.29}$}  & \multicolumn{2}{c}{$7.42^{+0.01}_{-1.93}$}  &   \multicolumn{2}{c}{$\mathcal J$(0.1,50)}   \\
    RV$_{\text{jit. HAPRS-pre}}$ (m s$^{-1}$) & \multicolumn{2}{c}{$8.39^{+0.08}_{-1.53}$}  & \multicolumn{2}{c}{$7.50^{+0.96}_{-0.64}$}  &   \multicolumn{2}{c}{$\mathcal J$(0.1,50)}   \\
    RV$_{\text{jit. HAPRS-post}}$ (m s$^{-1}$) & \multicolumn{2}{c}{$5.82^{+2.79}_{-0.13}$}   & \multicolumn{2}{c}{$8.44^{+0.22}_{-2.50}$}   &   \multicolumn{2}{c}{$\mathcal J$(0.1,50)}   \\  
    \hline
    \end{tabular*}
    \tablefoot{The adopted priors are listed in the right-most columns, with $\mathcal{U}$ and $\mathcal{J}$ denoting uniform and Jeffrey's priors, respectively.}
    \end{table*}

\subsection{Nested sampling}

\begin{figure*}
\centering
\includegraphics[width=0.9\hsize]{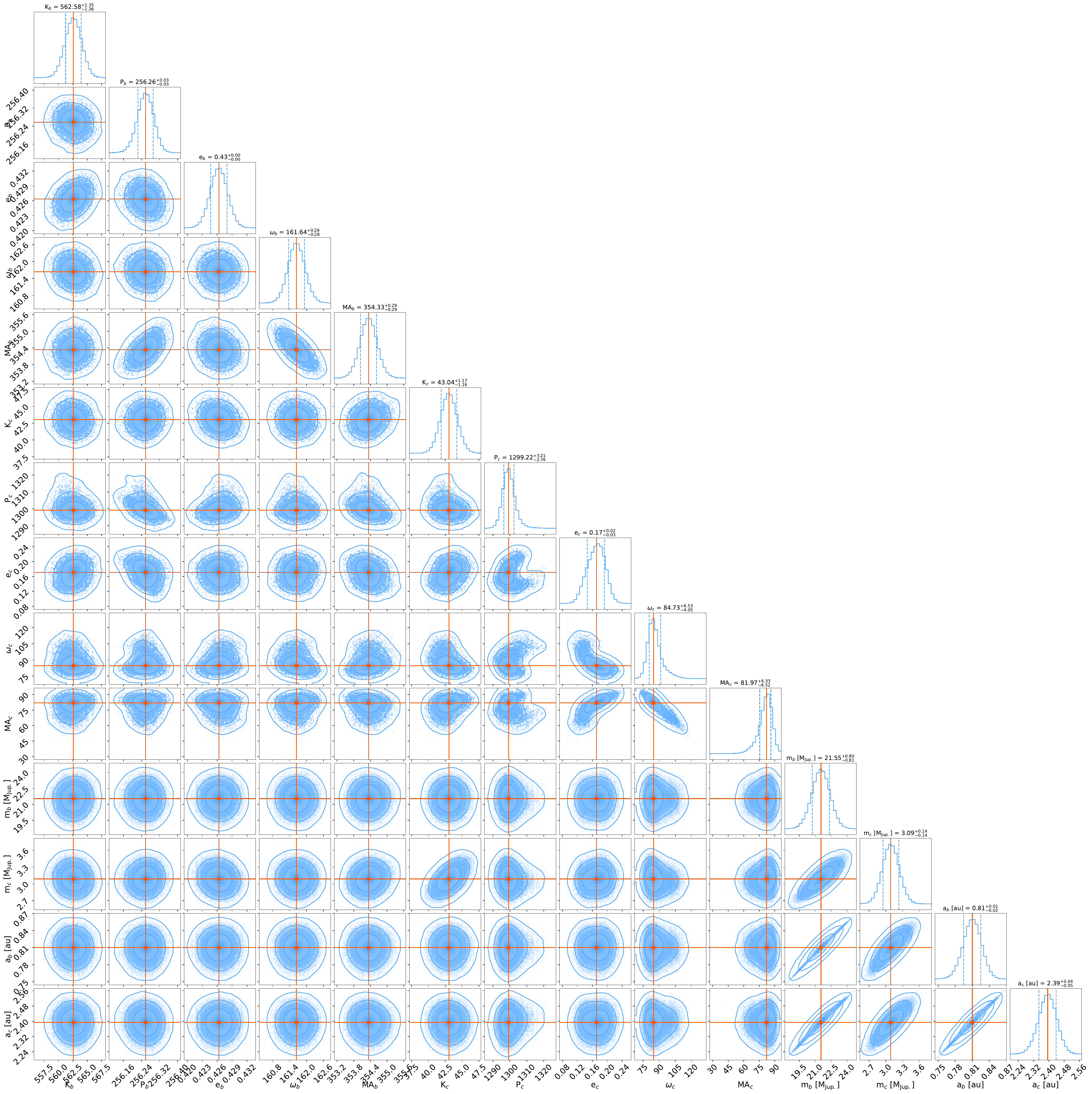}
\caption{Posterior probability distributions of the orbital parameters of HD 202206 from the nested sampling analysis of the coplanar inclined ($i$ = 51°) dynamical fit to the combined HARPS and CORALIE dataset. The solid red  lines indicate the median of each fitted parameter, and the 1$\sigma$ confidence intervals are indicated by the vertical dashed blue lines. The blue contours in the two-dimensional panels indicate the 1$\sigma$, 2$\sigma$, and 3$\sigma$ confidence intervals.}
    \label{NS}%
\end{figure*}

    We employed dynamic nested sampling (\citealt{Skilling2004}; \citealt{NS}) to perform posterior inference for the HD\,202206 system. We employed the dynamic nested sampling package \textsc{Dynesty} (\citealt{dynesty}), which explores the parameter space around the best-fit solution through random walks and provides a well-defined convergence criterion based on the remaining log-evidence, $\Delta\ln \mathcal{Z}$. This enables a robust assessment of orbital architecture and associated dynamical behaviors, thereby informing the long-term stability and plausible formation pathways of the system. Each fitting parameter was sampled with 100 live points, and the runs were terminated once the remaining log-evidence ($\Delta\ln \mathcal{Z}$) dropped below 0.001. These settings ensure a robust convergence of dynamic nested sampling. The median and $1\sigma$ range of the posteriors of the fitted parameters for the coplanar edge-on ($i = 90^\circ$) and inclined ($i = 51^\circ$) dynamical fits are listed in Table \ref{orbital param}, and the posterior probability distributions of the orbital parameters for the coplanar $i = 51^\circ$ dynamical fit are shown in Fig.~\ref{NS}.

\subsection{Orbital stability and libration of $\theta_4$}
\begin{figure}
\centering
\includegraphics[width=0.85\hsize]{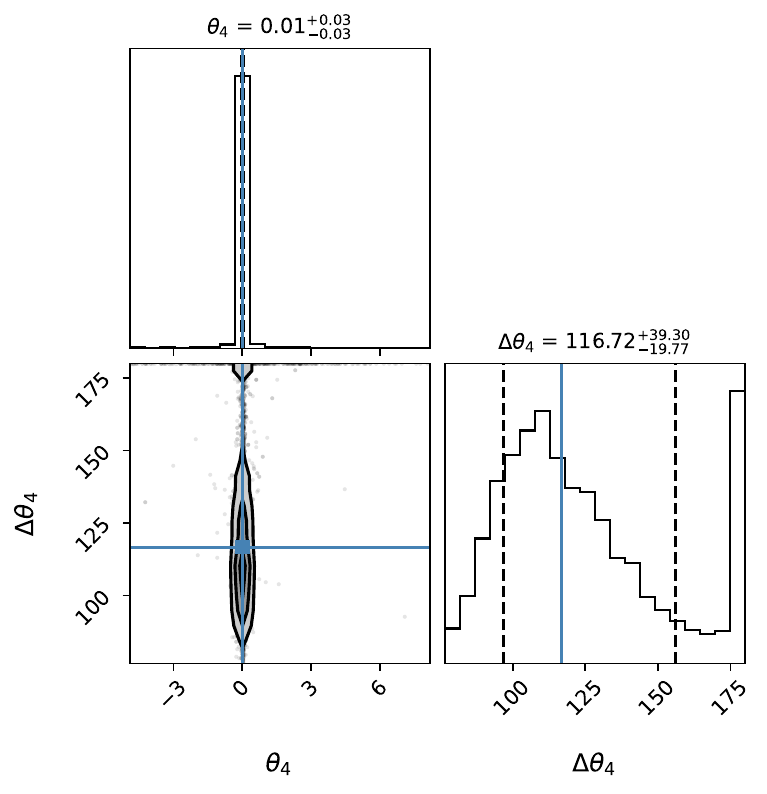}
    \caption{Distributions of the mean value of $\theta_4$ and the semi-amplitude, $\Delta\theta_4$, from a 10 Myr integration of 5000 subsamples extracted from the posterior of the coplanar inclined ($i$ = 51°) dynamical fit to the combined CORALIE and HARPS dataset of HD\,202206. The vertical solid blue   lines indicate the median of each parameter, and the vertical dashed black lines indicate the 1$\sigma$ confidence intervals. The black contours in the two-dimensional panels indicate the 1$\sigma$, 2$\sigma$, and 3$\sigma$ confidence intervals.}
    \label{conerplot_theta4}
\end{figure}

    We used $N$-body simulations performed with the Wisdom-Holman method \citep{Wisdom1991} up to a maximum of 10 Myr to analyze the statistics of long-term dynamics from 5000 subsamples randomly extracted from the nested sampling posterior distribution of the coplanar edge-on ($i = 90^\circ$) and inclined ($i = 51^\circ$) fits for the HD\,202206 system. A subsample is defined as unstable if $e > 0.9$ or the semimajor axis changes by more than 60\% from its initial value for either planet. Otherwise, it is considered stable. Thus, stability is the fraction of stable subsamples, while the libration rate of $\theta_4$ is the fraction of subsamples in which $\theta_4$ exhibits libration ($\Delta \theta_4 < 170^\circ$). For both inclinations, none of the subsamples show librations of $\theta_1$, $\theta_2$, $\theta_3$, or $\theta_5$. The results of this analysis are presented in Table~\ref{differentcases}. For $i = 90^\circ$, we find that 76.7\% of the subsamples exhibit a libration of $\theta_4$ about $0^\circ$. The libration ratio of $\theta_4$ increases to 88.1\% for $i = 51^\circ$. Figure~\ref{conerplot_theta4} presents the distributions of the mean value of $\theta_4$ and the semi-amplitude ($\Delta\theta_4$) about the mean for the $i=51^\circ$ subsamples. The librating cases show large amplitude libration, with the peak in the distribution of $\Delta\theta_4$ at $\sim 110^\circ$, while the circulating cases shift the median of $\Delta\theta_4$ to a slightly larger value of $116{\fdg}7$. It can be seen that the stable fraction is maintained above 99\% for both inclinations, indicating that dynamical stability does not provide an additional constraint on the libration or circulation of $\theta_4$. We also verified that the stable fraction and the libration ratio are not sensitive to the maximum integration time. They change by $\sim 1\%$ only when we integrate a smaller set of 2000 subsamples from the $i = 51^\circ$ posterior up to 100 Myr. Based on these findings, we conclude that the HD\,202206 system is in a broadly stable region of the parameter space and that it is most likely in the 5:1 MMR with only $\theta_4$ librating.

\begin{table}[t]
    \centering
    \caption{Stability and libration ratio of $\theta_4$ for different inclinations.}
    \renewcommand{\arraystretch}{1.5}
    \begin{tabular}{lcc}
    \hline\hline
         Inclination ($i$) & Stability (\%)  &  Libration Ratio of $\theta_4$ (\%)\\
    \hline
         90°& 99.96 &  76.73 \\
    \hline 
         51°& 99.78 &  88.10 \\
    \hline
    \end{tabular}
    \tablefoot{From 5000 subsamples integrated up to 10 Myr.}
    \label{differentcases}
\end{table}

\section{Discussion}
\label{sec7}
\subsection{Dominance of resonance angle $\theta_4$}

For a system with two companions b and c, which are on coplanar inner and outer orbits, respectively, the disturbing function for their gravitational interaction can be written as \citep{murray1999,Mardling2013}
\begin{equation}
\mathcal{R} = \sum_{m=0}^{\infty} \sum_{n'=-\infty}^{\infty} \sum_{n=-\infty}^{\infty} \mathcal{R}_{mn'n} \cos \phi_{mn'n} ,
\end{equation}
where the harmonic angle
\begin{equation}
\phi_{mn'n} = n'\lambda_b - n\lambda_c + (m-n')\varpi_b - (m-n)\varpi_c .
\end{equation}
For the $n$:$n'$ commensurability, the order of the resonance is $n - n'$, and there are $n-n'+1$ resonance angles ($\phi_{mn'n}$) with $m = 1, \ldots, n-n'+1$.
When the companions are far apart and the semimajor axis ratio ($\alpha = a_b/a_c$) is small, the strongest MMRs are the $n$:1 resonances with $n' = 1$ \citep{Gallardo2006, Mardling2008}.

In the HD 202206 system, we find that only the 5:1 resonance angle with $m = 2$ (i.e., $\phi_{2,1,5}= \theta_4 = \lambda_b - 5\lambda_c + \varpi_b + 3\varpi_c$) librates with a large amplitude. The remaining four 5:1 resonance angles circulate together with the apsidal angle $\Delta\varpi = \varpi_b - \varpi_c$.
The 6:1 MMR in the $\nu$ Oph system behaves differently, with all of its associated resonance angles exhibiting stable libration. However, the angle with $m = 2$ (i.e., $\phi_{2,1,6} = \lambda_b - 6\lambda_c + \varpi_b + 4\varpi_c$) also shows the smallest libration amplitude \citep{Quirrenbach2019}.
In addition, in his study of planetary orbits around the primary star of the HD~59686 binary system, \cite{Wongthesis2014} found that, for the orbits that are stabilized by $n$:$1$ MMRs with $n \sim 40$, the angle $\phi_{2,1,n}$ (with $m=2$) also exhibits the smallest libration amplitude.
This general behavior that the $m=2$ resonance angle is the dominant angle for $n$:$1$ MMRs requires an explanation.

Previous studies have shown, through an appropriate canonical transformation, that the first-order $n$:$(n-1)$ MMR problem can be reduced to an integrable one-degree-of-freedom problem with a single mixed resonance angle \citep{Sessin1984, Henrard1986, Wisdom1986, Batygin2013b, Nesvorny2016, Petit2020}. For higher-order resonances with $n/n' < 2$ or $P_c/P_b < 2$, \cite{Hadden2019} has shown that a near-symmetry of the disturbing function also allows the problem to be approximated as an integrable one-degree-of-freedom problem with a single mixed resonance angle. However, this reduction does not apply to $n$:$1$ resonances with $n > 2$.

For high-order $n$:1 MMRs where $n$ is large and $\alpha$ is small, an alternative approach is to expand the disturbing function in spherical harmonics or powers of $\alpha$ \citep{Mardling2013}:
\begin{equation}
\mathcal{R}_{mn'n} = \frac{G\mu_i m_c}{a_c}\sum_{l=l_{\min},2}^{\infty}\zeta_m c_{lm}^2\mathcal{M}_l \alpha^l X_{n'}^{l,m}(e_b) X_{n}^{-(l+1),m}(e_c) ,
\label{eq:R}
\end{equation}
where $X_{n'}^{l,m}(e_b)$ and $X_{n}^{-(l+1),m}(e_c)$ are Hansen coefficients and $l_{\min}=3$ for $m=1$ and $l_{\min}=m$ for $m \ge 2$ (see \citealt{Mardling2013} for the definitions of the other variables in Eq.\ \ref{eq:R}). For given $n$ and $n'$, only $R_{mn'n}$ with $m = 2$ includes a term with the lowest power of $\alpha$, which is the quadrupole term with $l = 2$ or $\alpha^2$. Other $R_{mn'n}$ with $m \not = 2$ include only terms of order $\alpha^3$ or higher. This explains why the $m=2$ resonance angle is usually the dominant one for high-order $n$:1 MMRs (see also \citealt{Mardling2008, Mardling2013}).

\subsection{Formation and migration scenario}

The existence of a MMR between two companions requires convergent migration and resonance capture \citep{Goldreich1965,murray1999}. High-order MMRs are generally difficult to assemble and maintain because the width of a resonance decreases with increasing resonance order \citep{murray1999, Mardling2008, Mardling2013}. Capture into high-order MMRs probably requires slow migration and nonzero initial eccentricities.

The 5:1 MMR in HD\,202206 and the 6:1 MMR in $\nu$ Oph are the only high-order $n$:1 resonances confirmed to date for systems with substellar companions beyond the Solar System. They are similar in having a brown dwarf inner companion, but they differ in other respects.
The $\nu$ Oph system involves two companions of comparable masses ($m_b \sin i = 22.2 M_\text{J}$ and $m_c \sin i = 24.7 M_\text{J}$) and eccentricities ($e_b = 0.12$ and $e_c = 0.18$), with all six resonance angles librating.
\cite{Quirrenbach2019} have argued that this configuration is most likely consistent with a scenario in which the two brown dwarfs formed and migrated in a disk around the star. The HD\,202206 system is a hierarchical system that hosts an inner brown dwarf with $m_b = 21.6 M_\text{J}$ and an outer giant planet with $m_c = 3.1 M_\text{J}$. There are two plausible formation scenarios: (1) the formation and migration of both the brown dwarf and the giant planet in a circumstellar disk 
(as proposed for $\nu$ Oph) or (2) the formation of the star--brown dwarf binary, followed by the formation and migration of the giant planet in a circumbinary disk.
In the context of the second scenario, it is interesting to note that simulations of planet migration in circumbinary disks around binary stars have shown capture into MMRs such as 4:1 and 5:1 \citep[e.g.,][]{Nelson2003, Kley2014, Zoppetti2018}. However, since none of the observed circumbinary planets are in a $n$:1 resonance with the binary, these works have proposed either a wider disk cavity to stop migration before MMR capture or subsequent evolution out of the MMR due to tides or scattering. Since the companions in HD\,202206 are in 5:1 MMRs, these additional processes must be ineffective, which could be due to the low mass and long period of the inner companion compared to the secondary stars of binaries with circumbinary planets.

The current dynamical architecture of the HD\,202206 system can be used to constrain the formation scenarios.
The large mass ratio $m_b/m_c = 6.9$ of the HD\,202206 companions means that it would be difficult to increase $e_b$ significantly after resonance capture, and the inner brown dwarf may be required to have a rather high eccentricity (comparable to its current value $e_b = 0.43$) before resonance capture. The large amplitude libration of only $\theta_4$ may be a consequence of the high $e_b$ before resonance capture.
A detailed exploration of the formation and migration scenarios of HD\,202206 and $\nu$ Oph, including the specific constraints on initial parameters and migration rates, will be presented in our forthcoming study (Cao et al., in prep).

\section{Summary}
\label{sec8}

We have presented a comprehensive dynamical analysis of the HD\,202206 system that we carried out by combining RV data from CORALIE and HARPS with astrometric constraints from \textit{Gaia} DR3 and \textsc{Hipparcos}. The inclusion of astrometric proper motion anomalies and jitter constraints resolved the previous ambiguity regarding the system orientation and constrained the orbital inclination to $50.7_{-1.4}^{+25.8}$ degrees. This result, as well as the poor quality of the dynamical fit to the RV data at low inclinations, allows us to rule out the low-inclination solution previously proposed by \cite{Benedict2017} and confirms that the inner companion, HD\,202206 b, possesses a true mass of $21.6 M_\text{J}$ in the brown dwarf regime, distinct from the planetary mass $3.1 M_\text{J}$ of the outer companion. The derived coplanar inclined ($i$ = $51^{\circ}$) solution yields orbital periods of 256.26 and 1299 days, with significant eccentricities of 0.426 and 0.180 for the inner and outer companions, respectively.

Our numerical $N$-body simulations confirm that the system is dynamically stable and is most likely in the 5:1 MMR. A detailed inspection of the resonance angles reveals that only the angle $\theta_4 = \lambda_b - 5\lambda_c + \varpi_b + 3\varpi_c$ is librating with a large amplitude; the other resonance angles circulate. The dominance of the angle involving one $\varpi_b$ for high-order $n$:1 MMRs is due to the fact that the quadrupole term in the disturbing function contributes only to the resonance term involving this angle.
 
The existence of a brown dwarf and a giant planet in such a high-order MMR strongly favors formation scenarios involving disk-induced migration. This could involve either the formation and migration of both companions in a circumstellar disk or the formation and migration of the giant planet in a circumbinary disk after the formation of the star--brown dwarf binary. The former scenario would imply that brown dwarfs can originate from planet-like formation channels within a circumstellar disk rather than solely through cloud fragmentation processes, as suggested previously for the $\nu$ Oph system, which consists of two brown-dwarf companions in a 6:1 MMR \citep{Quirrenbach2019}. The constraints on the formation history of HD\,202206 from its observed dynamical state will be investigated in detail in our future work.

\begin{acknowledgements}
Y.C.\ and M.H.L.\ were supported
in part by the Hong Kong RGC grant 17309323.
T. T.\ and S. S.\ acknowledge support by the Bulgarian National Science Fund BNSF program ``VIHREN--2021" project No. KP--06--DV/5/15.12.2021. 
\end{acknowledgements}

\bibliographystyle{aa}
\bibliography{references}

\end{document}